# High purity semi-insulating 4H–SiC epitaxial layers by Defect-Competition Epitaxy

MVS Chandrashekhar[1,a)], IftekharChowdhury[1], P. Kaminski[2], R. Kozlowski[2], P. B. Klein[3], Tangali Sudarshan[1]

[1]University of South Carolina, Electrical Engineering, Columbia, SC 29208

[2]Institute of Electronic Materials Technology, Warszawa, Poland

[3]Naval Research Laboratory, Washington D.C

## Abstract

Thick, high-purity semi-insulating (SI)homoepitaxial layers on Si-face 4H-SiC weregrownsystematically, with resistivity$\geq 10^9 \Omega$-cmby maintaining high C/Si ratios 1.3-15 during growth.Comparison of secondary ion mass spectra betweenlow-dopedepilayers grown at C/Si ratio<1.3andSI-epilayers grown at C/Si ratio>1.3 showed little difference in residual impurity concentrations. A reconciliation of impurity concentration with measured resistivity indicated a compensating trap concentration of $\sim 10^{15} cm^{-3}$present only in the SI-epilayers. High-resolution photo induced transient spectroscopy (HRPITS) identified themas Si-vacancy related deep centers, with no detectable $EH_{6/7}$ and $Z_{1/2}$levels. Recombination lifetimes ~5ns suggest application in fast-switching power devices.

---

[a)] Author to whom all correspondence should be addressed: Chandra.mvs@gmail.com

Silicon carbide (SiC) has the potential to replace conventional semiconductors in high frequency and high power applications such as in pulse-width modulated electric vehicles, smartgrids and next generation efficient power electronics. SiC's advantages over silicon, the current industry standard, include high saturation velocity (high current), wide bandgap (high voltages and temperature), both of which enable minimization of parasitic capacitances and reduction of active cooling, leading to transformational architectural improvements in power electronics.

While the device structures have been well optimized, other challenges in SiC technology remain, and revolve around material quality. Research in the past few years has focused on growing high-quality single crystal SiC substrates and epilayers with low densities of structural defects. In addition to the high structural quality there are other requirements that must be met for these epilayers to be useful in high frequency and high power devices. In a power transistor chip the metalized backside and fingers on the epitaxial side introduce a passive parasitic capacitance, along with the active capacitive part of the device, limiting the performance at high frequencies. This parasitic capacitance can be minimized by using semi-insulating (SI) epilayers/substrates. One of the methods for introducing SI property to the substrate is to use vanadium as a deep level dopant to pin the Fermi level near mid bandgap. Although this is the original conceived method for producing commercial SI substrates, reports[1]indicate that vanadium degrades crystal quality supported by increased FWHM of X-ray diffraction rocking curves.

The alternative is to pin the Fermi level deep in the bandgapby compensating shallow donor and acceptor levels from residual impurities with intrinsic deep level defects. The development of such high purity semi-insulating (HPSI) SiC was reported[2] in 2000, but determination of the exact nature of the intrinsic defects and the compensation mechanism is still in progress. EPR measurements on as-grown HPSI 4H-SiC detected[3, 4] carbon vacancies $V_C$.A broad range of defects including $V_{Si}$, $V_C$-$V_{Si}$ and $V_C$-$C_{Si}$ have been reported by N T Son[5]. A recent report from Mitchelet al[6]summarized the deep levels present in the upper half of the band gap of 4H-SiC HPSI material which includes $Z_{1/2}$, $RD_{1/2}$ and $EH_{6/7}$. $Z_{1/2}$ is an efficient recombination center and act as lifetime killing defects[7] in 4H-SiC. Identification of the electronic levels due to intrinsic defects responsible for lifetime control and compensation in HPSI 4H-SiC is of great technological importance.

In this letter we discuss the chemical vapor deposition (CVD) growth conditions for achieving HPSI epitaxial layer by defect competition epitaxy and identify the deep defect centers, responsible for SI behavior, using high resolution photo induced transient spectroscopy (HRPITS). The epitaxial films discussed here were all grown on commercially available 4H-SiC, n-type Si-face substrates (Cree Inc.) polished $8^0$ off-axis from the (0001) plane towards (112-0) plane, in a home-built vertical hot-wall CVD reactor at temperatures $1450^0$-$1650^0$ C and pressure 80-120 torr. Dichlorosilane (Si source) and propane (C source) are the main precursor gases with hydrogen as carrier gas. The growth rate was kept at 30 μm/hr and no intentional dopant was incorporated during growth. Further details are discussed elsewhere[8].The epilayers were characterized using X-ray diffraction (XRD) rocking curve for crystal quality, micro-Raman spectroscopy for polytype uniformity, transmission line model (TLM) for resistivity, time resolved photoluminescence spectroscopy (TRPL) for carrier lifetime, secondary ion mass spectroscopy (SIMS) for impurity concentration and high resolution photo induced transient spectroscopy (HRPITS) for characterizing trap centers. SIMS was performed by Evans Analytical Group (EAG), NJ.

Site competition epitaxy[9]allows for controlling background dopant incorporation into SiC by adjusting the C/Si ratio. Nitrogen (n-type dopant) competes for C sites and aluminum (p-type dopant) competes for Si sites of the growing SiCepilayer. Increasing C/Si ratio decreases nitrogen incorporation, making the epilayer less n-type, while decreasing C/Si ratio, decreases aluminum incorporation, making the epilayer less p-type. It is expected that there is a transition region between n and p-type where impurity incorporation is minimized to produce SI material when the necessary deep levels are present. This was the rationale behind varying C/Si ratio to produce HPSI material.

A wide variation of epilayer doping with respect to C/Si ratio was observed, as in Fig. 1. For 1.3<C/Si<1.5, the as grown epilayer shows SI behavior. We term this regime "defect competition epitaxy" to distinguish it from the "site competition" regime shown in the figure for reasons that will be discussed further below. Approximately twenty experiments were done over this defect competition regime.They displayed resistivity as high as $\rho > 10^9$ Ω-cm supported by TLM measurements. From TLM, a resistivity of $\rho = 1.6\times10^9$ Ω-cm was measured for SI samples with C/Si~1.4. We note that this is a lower bound, as the TLM patterns were not mesa isolated due to the thickness of the films (60μm), and therefore the spreading resistance is not accounted for. A typical epilayer X-ray rocking curve FWHM is ~8 arcsec (Fig. 2), close to the FWHM 6-8 arcsec of perfect SiC crystals[10]. This compares very favorably with recent results[11] showing FWHM of 18 arcsec for HPSI and 24 arcsec for V-doped SI.

To determine how C/Si affects background impurity incorporation, we have performed SIMS analysis. Table I shows N, Al & B concentrations in low doped n-type (C/Si~0.9) and in HPSI (C/Si~1.4) samples, along with the best published HPSISIMS result[12] to date. Surprisingly, we see that increasing C/Si from 0.9 to 1.4 does not significantly change the concentration of impurity incorporation but does change the resistivity from ~30Ω-cm to ~$10^9$Ω-cm. Hence, at these low background impurity levels, rather than driving out nitrogen, higher C/Si introduces intrinsic deep levels[13]which account for pinning the Fermi level far away from the band edge. High C/Si ratio promotes Si-vacancy ($V_{Si}$) and low C/Si ratio promotes C-vacancy ($V_C$) formation[13]. The variation of these intrinsic defects with C/Si ratio, as opposed to impurity incorporation, leads to the term defect-competition, in contrast with site-competition.Vanadium compensation can be excluded completely, as it is below the SIMS detection limit (< $2x10^{12}$ at/$cm^3$).

Using the donor and acceptor concentration from SIMS analysis (Table I), together with the doping and resistivity data, the schematic band diagram using Fermi statistics for an extrinsic semiconductor[14] can be determined. Figure 3(a) shows the band diagram for low doped n-type (C/Si: 0.9, doping from C-V= 1.5x $10^{15}$ $cm^{-3}$) sample, where the Fermi level ($E_f$)is 0.25 eV below the conduction band ($E_c$). Here we have taken the band gap Eg ≈ 3.26 eV, $E_c$-$E_d$≈ 90 meV[15], $E_a$-$E_v$≈ 220 meV[15] and $n_o$= ($N_d$-$N_a$) =$1.45\times10^{15}$ $cm^{-3}$.

The schematic band diagram for HPSI sample (Figure 4(a)) is determined using: resistivity $\rho = 1.6\times10^9$ Ω-cm, mobility $= 900$ $cm^{-2}$/v.s[16]for 4H-SiC, and the equation, $\rho = \frac{1}{nq}$ resulting in free electron density, $n \approx 10^6 cm^{-3}$. From the equation, $n = n_i \exp\left(\frac{E_F - E_i}{kT}\right)$ we see that for every decade change of free electron density, the Fermi level changes by 0.06 eV[14]. Given that the SiC effective density of states in the conduction band $N_c=1.5x10^{19}cm^{-3}$[17], $E_f$ for HPSI (C/Si~1.4) is located 0.8 eV below $E_c$. Though the actual mobility of SI material might be little lower, this value is taken as a template in the absence of a proper SI mobility value. We have to note that taking the mobility any values in the range of $100\text{-}1000cm^{-2}/v.s$ still pins the fermi level around 0.7-0.8 eV below $E_c$. This $E_f$ position is not consistent with the SIMS data unless the residual donors ($(N_d\text{-}N_a) = 1.45\times10^{15}\,cm^{-3}$) are completely compensated by intrinsic deep defects pinning $E_f$ at 0.8eV below $E_c$. From this analysis, the intrinsic trap concentration must be > $1.45\times10^{15}\,cm^{-3}$.

To confirm the presence of the inferred traps, TRPL measurements were performed on both samples at room temperature. Excitation was provided by a frequency doubled, modelocked, cavity-dumped Ti: sapphire laser (355 nm, 150 fsec pulse width, 100-500 kHz, 5nJ/pulse. The average injection level over the layer was $\approx 2x10^{14}cm^{-3}$. The n-type sample showed a lifetime of 0.8 μs[8] whereas the HPSI sample showed no epilayer band edge emission. This quenched band edge emission occurs if either the crystal quality is poor, or if there are enough traps to provide effective non-radiative recombination centers. The XRD FWHM ~8 arcsec (Figure 2) demonstrates high quality of these crystals, while SIMS demonstrates their high purity, which indicates that the short lifetimes suggested by the quenched band edge emission in the HPSI sample is from intrinsic deep defects.

To identify the nature of these intrinsic defects HRPITS measurements were performed on both low doped n-type and SI samples. HRPITS is a characterization technique in which the sample is illuminated with pulsed light. The resulting photocurrent transients are recorded as a function of temperature (77-750K), revealing information about traps in the material. The experimental details and photocurrent relaxation analysis are found elsewhere[18]. HRPITS in low doped n-type (C/Si: 0.9) samples revealed[19] [Figure 3(b)] the presence of $EH_{6/7}$, ($V_C$ related traps), as expected[13] under Si-rich growth conditions. The common shallow impurity level boron was not observed, in agreement with SIMS data (Table I).

HRPITS on HPSI (C/Si: 1.4) sample reveals[20] [Figure 4(b)] a contrasting picture. Only Si-vacancy, $V_{Si}$ and Carbon-antisite, $C_{Si}$ related deep defect levels were observed, in accordance with the increased C/Si ratio, where $V_{Si}$ and $C_{Si}$ are expected to dominate [13]. The positions of these levels cluster around 0.8eV below $E_c$, in agreement with the trap-pinned $E_f$ position extracted from TLM above. Also we see that the dominance of residual shallow donors (SIMS in Table I) favors negatively ($V_{Si}^{-}$) charged defects[13]. $Z_{1/2}$ and $EH_{6/7}$ levels are not observed, again in accordance with the increased C/Si ratio, which suppresses the formation of these levels[21]. It is important to note that from SIMS (Table I) we exclude the possibility of vanadium incorporation in our HPSI epilayer. In other words, by adjusting C/Si, the dominant generation/recombination (G-R) energy level has been moved from near mid-gap[22] (typical for SiC) to far away (» 3kT) from mid-gap.

This non-mid gap Fermi level in this HPSI SiC has important implications in high frequency SiC power switching devices. Using Shockley-Reed-Hall (SRH) recombination statistics[14], the recombination lifetime

$$\tau_r \equiv \frac{1 + \left( \frac{2n_i}{n_{no} + p_{no}} \right) Cosh\left( \frac{E_t - E_i}{kT} \right)}{v_{th} \sigma_o N_t} \quad (1)$$

is inversely proportional to $N_t$ , the concentration of recombination centers/traps; where $n_i$, $v_{th}$, $\sigma_o$,and$E_t$ are intrinsic carrier concentration, thermal velocity, capture cross section, and trap level respectively. For high speed switching operations, short recombination lifetimes are required. Deeper levels provide faster recombination. Using $N_t$ ~$1x10^{16}$ $cm^{-3}$, $\sigma_o = 10^{-15}$ $cm^2$[14], $n_i = 4.7x10^{-9}$ $cm^{-3}$[16]and $n_{no}= 10^6$ $cm^{-3}$ we estimate $\tau_r$ = 7 ns for the HPSI material, compared with 5ns for midgap G-R centers (Table II). This opens the possibility of switching speeds in the gigahertz range with the HPSI. In other words, by shifting the dominant G-R level away from midgap, switching speed can be maintained (Table II). However, the carrier generation time, which controls leakage current in bipolar devices[14], given by

$$\frac{\tau_g}{\tau_r} \cong 2Cosh\left( \frac{E_t - E_i}{kT} \right) \quad (2)$$

increases. In other words, $\tau_g$ becomes longer as $E_t$ moves away from the mid bandgap, strongly suppressing leakage current (Table II), which in SiC is dominated by generation current, $J_{gen} \cong \frac{qn_iW}{\tau_g}$. This may enable much higher blocking voltages than those achievable for devices with mid-gap G-R centers[22]. Here while we assume that the same levels are responsible for generation and recombination to simplify this discussion, we note that this may not necessarily be the case and detailed dynamical calculations must be performed to numerically determine effective recombination rates. Nevertheless, this does not change the conclusion that moving the major traps away from mid-gap suppresses leakage current; leading to larger blocking voltages.The demonstrated HPSI material with $V_{si}$-related trap levels at 0.8 eV may dramatically increase the blocking voltage in SiC fast-switching bipolar devices.

In summary, thick HPSI 4H-SiC (0001) epilayers were grown systematically in a vertical hot-wall CVD reactor at a high growth rate of 30 μm/hr. A wide window of defect competition epitaxy has been demonstrated. Resistivity $\geq 10^9$ Ω-cm has been shown by TLM. High C/Si ratio is the key for introducing semi-insulating behavior, supported by the presence of only Si-vacancy related defects shown by HRPITS measurements. These HPSI materials with G-R centers away from mid-gap can be used in fast-switching bipolar devices with high blocking voltage.

The authors thank contract monitor Dr. H. Scott Coombe of ONR for his support (Grant #N000140910619) of this research, and Drs. Kurt Gaskill and Rachael Myers-Ward at the Naval Research Laboratory for their assistance with the SIMS measurements.

TABLE I. SIMS analysis of 4H-SiC HPSI and n-type sample. The only elements detected are the shallow level impurities nitrogen and boron.

| SIMS data comparison | | | |
|---|---|---|---|
| Elements | n-type(C/Si:0.9) | HPSI(C/Si:1.4) | HPSI Cree[a)] |
| B | 1.0E+14 | 3.6E+14 | 7.0E+15 |
| N | 1.6E+15 | 1.8E+15 | 2.0E+16 |
| Al | 5.0E+13 | <7.0E+13 | <5.0E+13 |
| V | | <2.0E+12 | 5.0E+13 |
| Resistivity (Ω-cm) | ~30 | ~1.0E9 | ~1.0E11 |
| a) Ref. 12 | | | |

**TABLE I**

TABLE II. Comparison of calculated recombination and generation lifetimes for mid bandgap and $V_{si}$ related trap levels and their device implication.

| | Midbandgap<br>$E_t = E_c$-1.63 eV | $V_{si}$relatedtraps<br>$E_t = E_c$-0.8 eV | Deviceimplication |
|---|---|---|---|
| $\tau_r$ | 5 ns | 7 ns | Fast switching |
| $\tau_g$ | $2\tau_r$ | $8 \times 10^{13} \times \tau_r$ | Lowleakagecurrent |

**TABLE II**

**Figure Captions**

Figure 1: C/Si ratio dependence of the doping concentration for the unintentionally doped epilayers. The red and blue lines are trend lines to guide the eye to distinguish n-type from p-type behavior, respectively.

Figure 2: High resolution X-ray diffraction rocking curve of a 60 μm thick epilayer grown on a 4H-SiC, $8^0$ off towards (11-20) substrate

Figure 3: (a)Schematic band diagram for n-type (C/Si: 0.9) (b) Deep defect levels located using HRPITS

Figure 4: (a) Schematic band diagram for HPSI (C/Si: 1.4) (b) Deep defect levels located using HRPITS

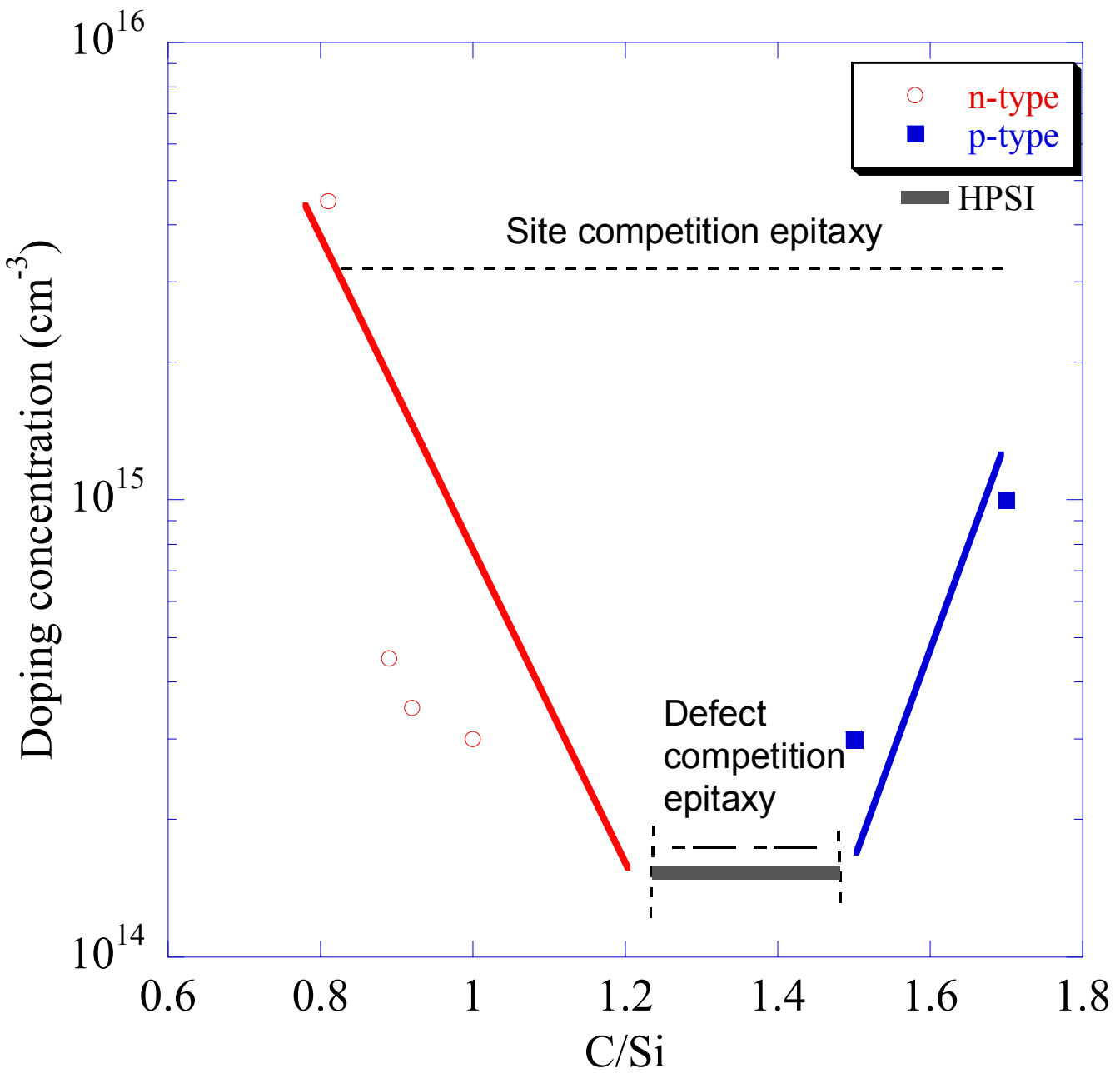

Figure 1

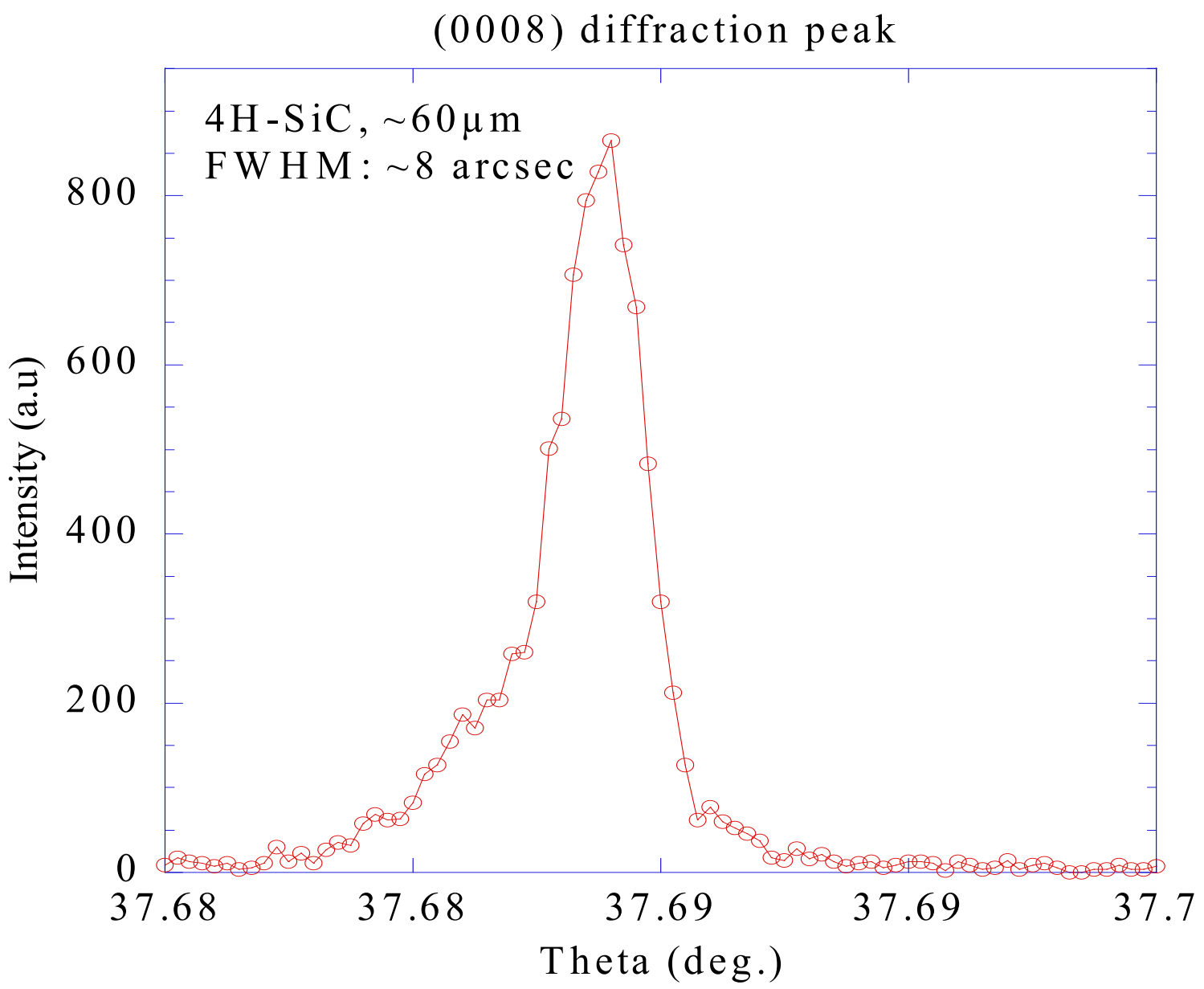

Figure 2

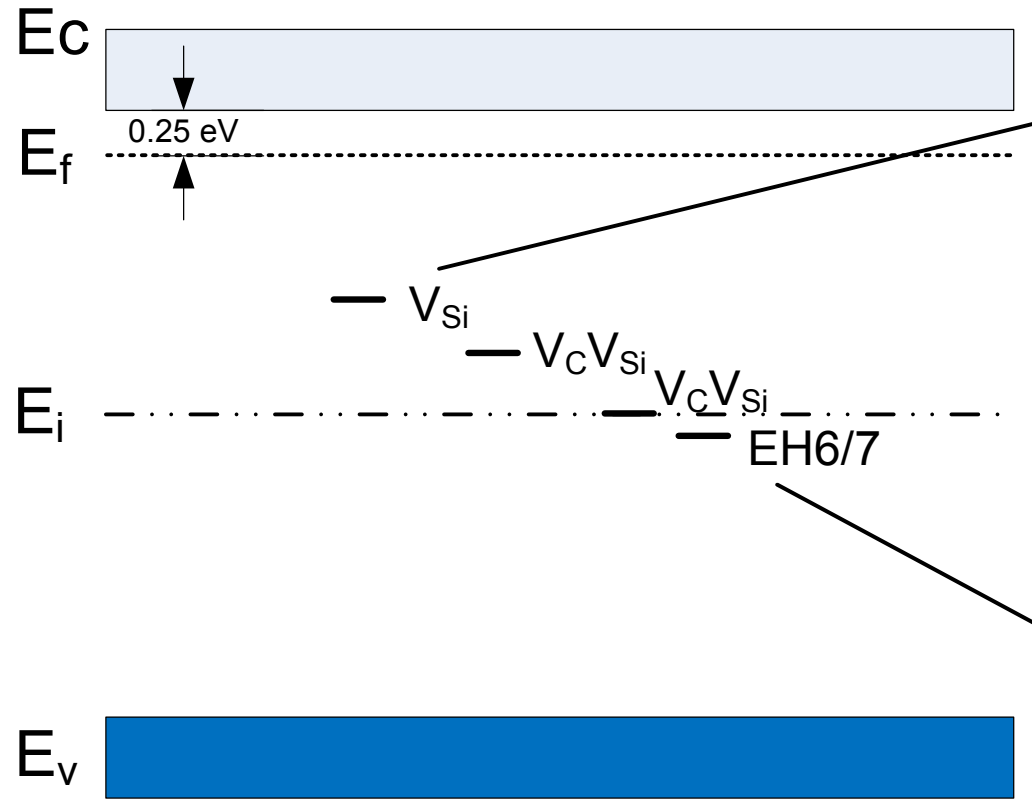

| $E_a$ * (meV) | Tentative identification |
|---|---|
| 975±30 | $V_{Si}^{2-/3-}$ a) |
| 1280±30 | $V_C C_{Si}^{0/-}$ a) |
| 1570±30 | $V_C V_{Si}^{0/-}$ a), *h* site |
| 1680±30 | *EH6/7* ($E_c$-1650 meV) a), $V_C V_{Si}^{0/-}$ *k* site |
| a) Ref. 19 | |

* $E_a$ : activation energy

(a) (b)

Figure 3

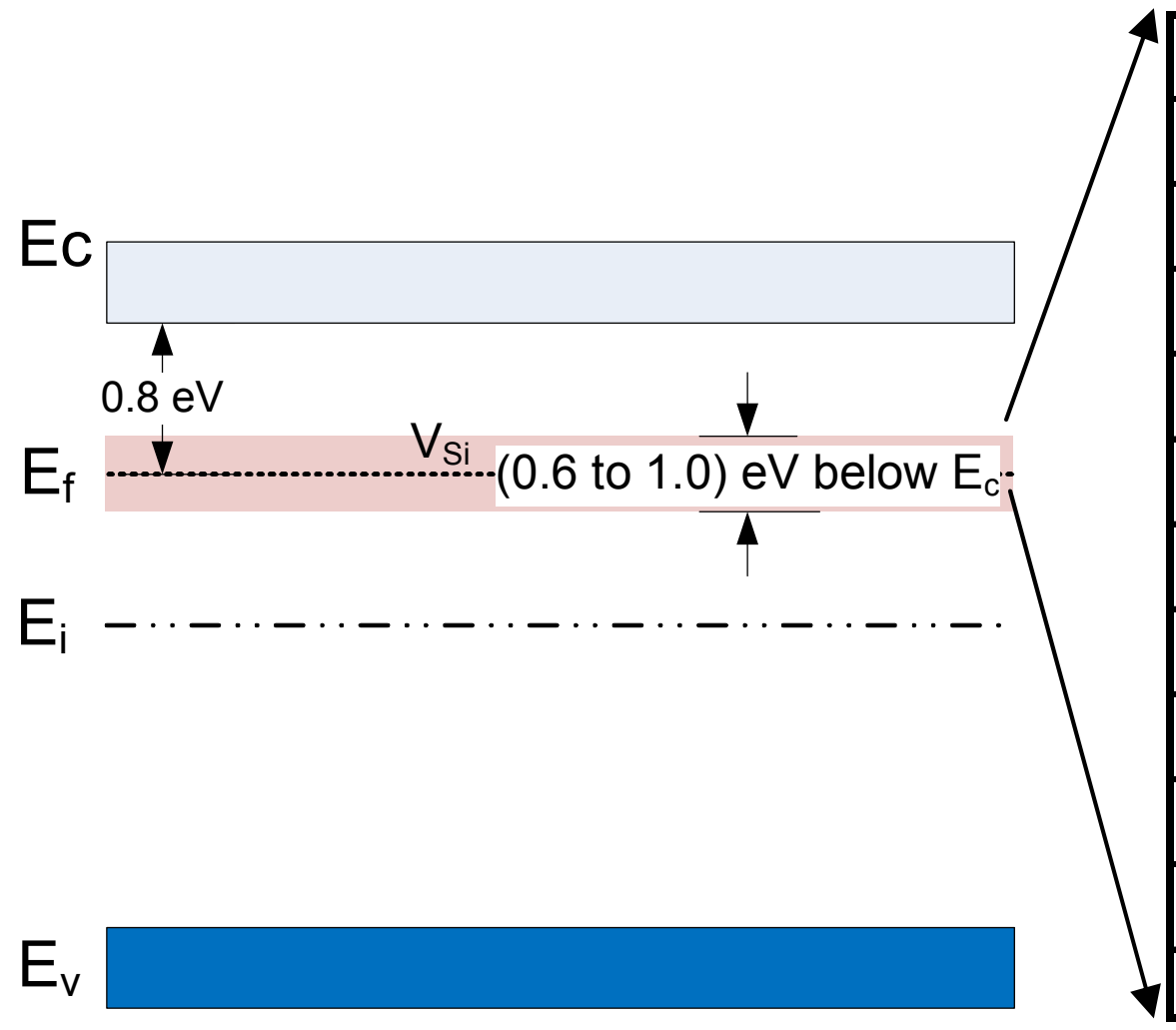

| $E_a$* (meV) | Tentative identification |
|---|---|
| 620±30 | $V_{Si}^{3-/4-}$ [a)], *h* site (*EH2* level ) [b)] |
| 650±30 | $V_{Si}^{3-/4-}$ [a)], *k* site (EH2 level ) [b)] |
| 740±30 | $V_{Si}^{2-/3-}$ [a)], *h* site |
| 770±30 | $V_{Si}^{2-/3-}$ [a)], *k* site |
| 820±30 | $V_{Si}^{1-/2-}$ [a)], *h* site |
| 890±30 | $V_{Si}^{1-/2-}$ [a)], *k* site |
| 930±30 | $V_{Si}^{0/1-}$ [a)], *h* site |
| 985±30 | $V_{Si}^{0/1-}$ [a)], *k* site |
| 1035±30 | $V_C C_{Si}^{1-/2-}$ [a)], *h* site |
| 1095±30 | $V_C C_{Si}^{1-/2-}$ [a)], *k* site |
| a) Ref. 19 b) Ref. 20 | |

(a)(b) * $E_a$ : activation energy

Figure 4